\documentclass[a4paper,twocolumn]{esapub} 

\usepackage{natbib}

\usepackage{graphicx}

\title{Scientific Performance of the ISDC Quick Look Analysis}

\author[2,1]{~~~~~~~~S. E. Shaw}

\affil[1]{INTEGRAL Science Data Centre, 16 Chemin d'\'Ecogia, CH-1290 Versoix, Switzerland~~~~~~~~~~~~~~~~~~~~~~~~~~~~~}

\affil[2]{School of Physics and Astronomy, University of Southampton, Hampshire, SO17 1BJ, UK~~~~~~~~~~~~~~~~~~~~~~~~}

\author[1,4]{N. Mowlavi}

\author[1,8]{K. Ebisawa}

\author[1,9]{A. Paizis}

\author[3,1]{J. Rodriguez}

\affil[3]{Service d'Astrophysique, Centre d'Etudes de Saclay, 91190 Gif-sur-Yvette, Cedex, France~~~~~~~~~~~~~~~~~~~~~~~~~} 

\author[1,4]{\newline J. Zurita}

\author[1,4]{R. Walter}

\affil[4]{Observatoire de Gen\`eve, 51 Chemin des Maillettes, CH-1290 Sauverny, Switzerland~~~~~~~~~~~~~~~~~~~~~~~~~~~~~~~~~}

\author[1,4]{M. T\"urler}

\author[1]{J. Soldan}

\author[3]{A. Sauvageon}

\author[1]{N. Produit}

\author[5,1]{\newline K. Pottschmidt}

\affil[5]{Max-Planck-Institut f\"ur Extraterrestrische Physik, Postfach 1312, 85748 Garching, Germany~~~~~~~~~~~~~~~~~~~~}

\author[1]{P. Meynis}

\author[1]{L. Martins}

\author[5]{L. Lerusse}

\author[6,1]{I. Kreykenbohm}

\affil[6]{Institut f\"ur Astronomie und Astrophysik T\"ubingen, Sand 1, 72076 T\"ubingen, Germany~~~~~~~~~~~~~~~~~~~~~~~~~~~}

\author[5,1]{\newline P. Kretschmar}

\author[1]{P. Haymoz}

\author[1,4]{P. Favre}

\author[1,4]{P. Dubath}

\author[1,4]{S. Deluit}

\author[1,4]{\newline T.J.-L. Courvoisier}

\author[7,1]{M. Chernyakova}

\affil[7]{Astro Space Centre, PN Lebedev Physical Institute, 84/32 Profsoyuznaya Street, Moscow 117997, Russia~~~~}

\author[1,4]{A. Bodaghee}

\author[8]{V. Beckmann}

\affil[8]{NASA Goddard Space Flight Center, Code 661, Building 2, Greenbelt, MD 20771, USA~~~~~~~~~~~~~~~~~~~~~~~~~~}

\affil[9]{CNR-IASF, Sezione di Milano, Via Bassini 15, 20133 Milano, Italy~~~~~~~~~~~~~~~~~~~~~~~~~~~~~~~~~~~~~~~~~~~~~~~~~~~~~~}

\begin{document}


\maketitle

\begin{abstract}

The INTEGRAL Science Data Centre (ISDC) routinely monitors the Near Real Time data (NRT) from the INTEGRAL satellite.  A first scientific analysis is made in order to check for the detection of new, transient or highly  variable sources in the data.  Of primary importance for this work is the Interactive Quick Look Analysis (IQLA), which produces JEM-X and ISGRI images and monitors them for  interesting astrophysical events.

\end{abstract}

\section{Introduction}

The Operations Team of the INTEGRAL Science Data Centre (ISDC) provides coverage for the mission every day of the year, between the hours of 9am and 6pm, and consists of a Coordinator, two Operators and a Scientist on Duty drawn from a pool of ISDC collaborators \citep{isdc}.  Near Real Time (NRT) telemetry is received continuously from the satellite, via the Mission Operations Centre (MOC), at a rate of $\sim100$ kbits/s \citep{ground}.  Of the total mission time, approximately 78\% is spent in science pointings.  The Operations Team is responsible for checking, processing and archiving the telemetry data, as well as distributing data products to the scientific community with the archive team.

Within the Operations Team, the Scientist on Duty is primarily responsible for making the first scientific analysis of the data, monitoring the status of the scientific instruments, and also dealing with triggers from the INTEGRAL Burst Alert System \citep{ibas}.  The Interactive Quick Look Analysis (IQLA) software is used to display high energy images, as soon as they are available, and allows the ISDC Scientist on Duty to inspect each image for interesting astrophysical events and to confirm alerts raised by the analysis pipeline.

\begin{figure*}[ht]

\centering

\includegraphics[width=0.8\linewidth]{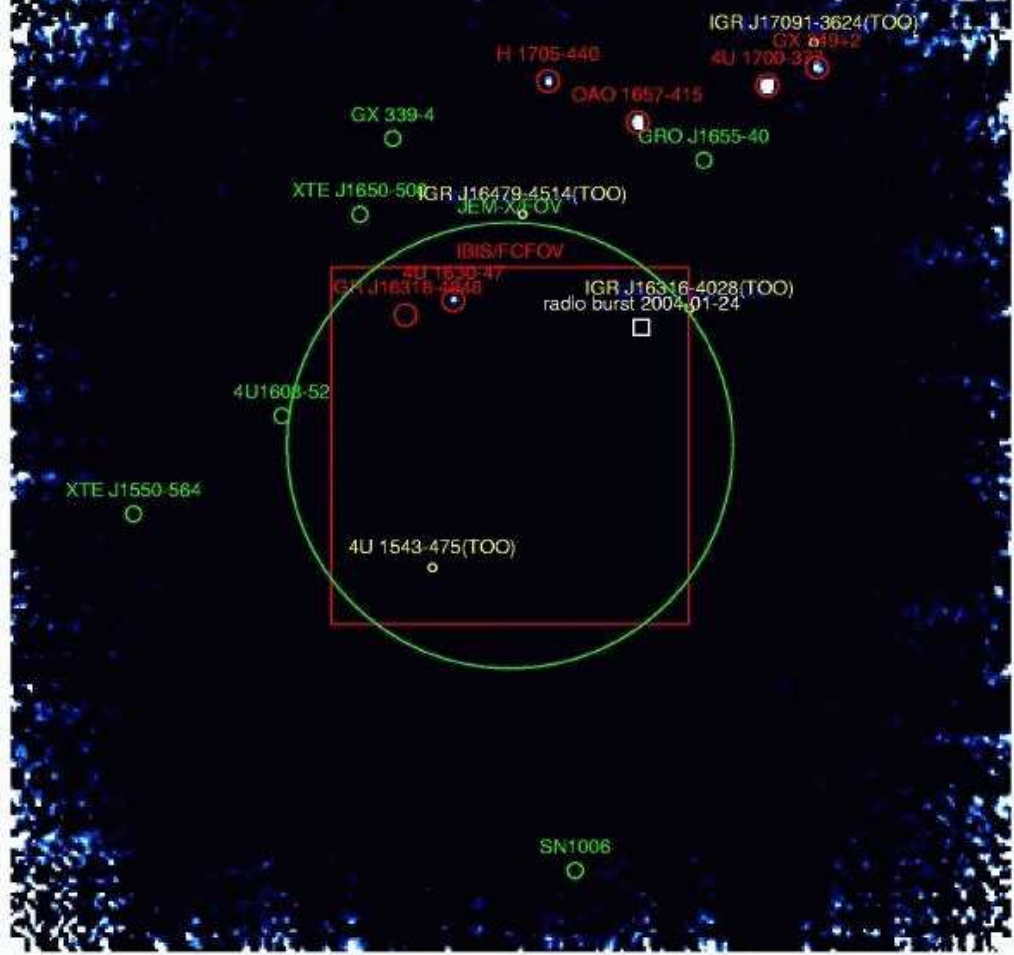}

\caption{A typical IQLA image, made from ISGRI 20--60 keV NRT data.  A number of sources are clearly detected.  The Scientist on Duty can fully interact with the display and show various information, such as:  Catalogue sources that are detected in the image by IQLA; Undetected catalogue sources above a user specified counts threshold; Sources not in the catalogue for which an open program proposal has been accepted; New sources or transients that are not in the current catalogue release;  Other sources of interest.\label{fig1}}

\end{figure*}

\section{ISDC Science Monitoring}

The NRT telemetry is received by the Operators and processed by a series of autonomous analysis `pipelines' running on the operations computer network, which is isolated from the outside world.  Images for the JEM-X \citep{jemx} and ISGRI \citep{isgri} instruments are automatically produced for each Science Window (ScW), with a delay of about 2 hours from the end of the ScW.  The length of one ScW is generally between 0.5--1.0 hours long, depending on the specific observation parameters \citep{performance}.  Approximately 2\% of the available science data is lost in short gaps, which do not affect IQLA.  However, IQLA is not able to produce images for a further $< 2\%$ of data where the gaps have a significant duration when compared to the length of the ScW.

Images are made in two bands for both instruments: 3--10 keV and
10--30 keV for JEM-X, and 20--60 keV and 60--200 keV for ISGRI.
Images are not made for SPI since it has poorer sensitivity and
angular resolution than ISGRI on the ScW time scale.  The ISDC analysis pipeline issues alerts for potentially interesting astrophysical events, using the ISDC General Reference Catalogue of high energy sources as an input \citep{cat}.  A detection is identified with a particular source, for both JEM-X and ISGRI, if it is located within 0.1$^{\circ}$ of the catalogue position.  Alerts are then issued in three cases:

\begin{itemize}

\item  \emph{NEW}:  A source is detected with a significance above a certain threshold ($> 16\sigma$ for JEM-X, $> 10\sigma$ for ISGRI) and can not be associated with any known position from the ISDC reference catalogue.

\item  \emph{TOO}:  A source, which is flagged as having an accepted INTEGRAL Target of Opportunity (TOO) proposal, is detected above $16\sigma$ in JEM-X or $10\sigma$ in ISGRI.  The Scientist on Duty then performs further analysis to check if the trigger criterea specific to that source are met, before taking further action.

\item  \emph{VAR}:  A source is detected and its brightness is measured to vary, by  $\pm$ a factor of 10, of a baseline flux recorded in the catalogue for each instrument and energy band.

\end{itemize}

Very bright \emph{NEW} source alerts, with detection significance $>
30\sigma$ in either JEM-X or ISGRI, are immediately sent by automated SMS text message to the ISDC Operations Coordinator's mobile telephone, which
enables an instant response at any time of the day.  The
values of the alert thresholds depend on the performance of the imaging software, which is affected by
systematic errors.  The present values have been set, to minimise the number of false alerts
while not missing important events, by experience from
the first year of operations.  

\begin{figure*}

\centering
\includegraphics[width=0.8\linewidth]{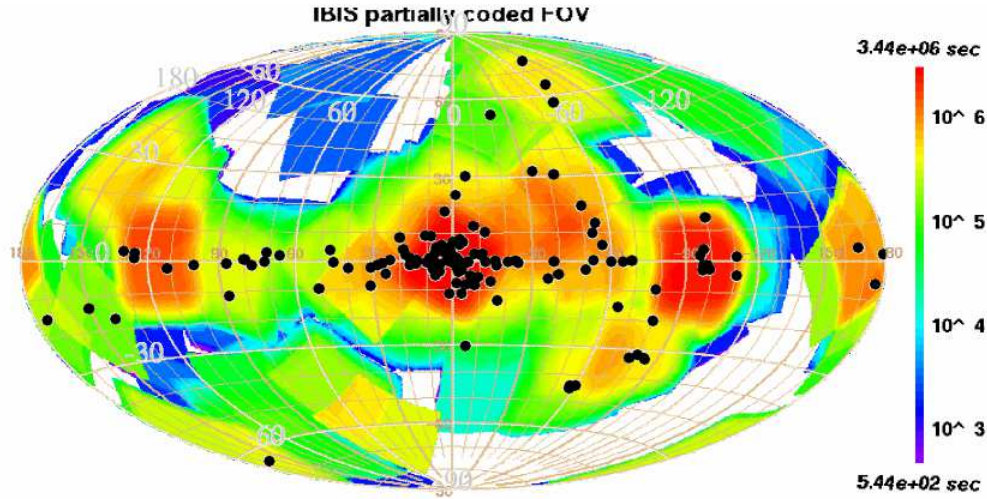} 
\caption{The ISGRI exposure map, in Galactic Coordinates, for the period 2003-06-12 to 2004-02-10 (Revolutions 81 to 162).  Superimposed are the positions of 157 sources detected in the same period by IQLA in one energy band and at least one ScW with either JEM-X or ISGRI.\label{exp}}

\end{figure*}

In addition, the  IQLA software scans each image and produces a summary of the sources detected in it, containing the detected position and flux of each source together with expected values from the catalogue, where appropriate.  The science output from IQLA is monitored in real time, during the working day, 7 days a week by  the ISDC duty scientist.  Images taken during the night are inspected routinely each morning.  In particular, the images  are examined when an automatic alert has been issued.  A typical IQLA ISGRI image, of a single ScW, from recent GPS observations is seen in Fig.~\ref{fig1}.

\section{Source Histories}

\begin{figure*}

\centering

\includegraphics[width=0.8\linewidth]{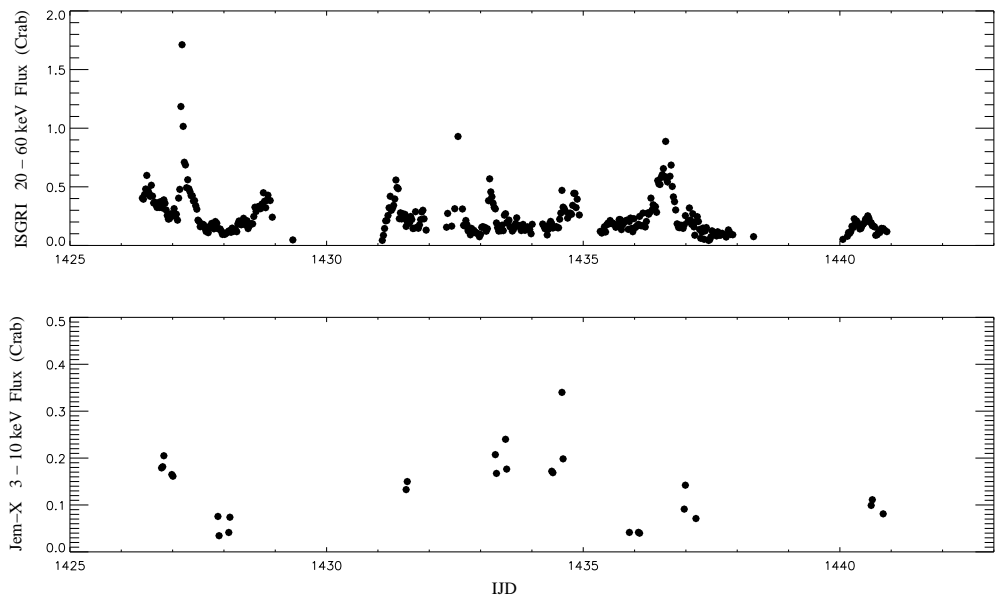}

\caption{Light curves of Vela X-1, taken from ISGRI
  and JEM-X IQLA images, with fluxes scaled to the Crab Nebula.
  A bright flare can be seen, starting at 2003-11-28T02:32:52 UTC
  (INTEGRAL Julian Date $= 1427.1069$).  The relative sparcity of the JEM-X
  data is due to that instrument's smaller
  field of view and the 5$\times$5 dither pattern of the observation.
  See also Staubert (in these proceedings) for analysis of this event.\label{vela}}

\end{figure*}

Since Revolution 81 (2003-06-12) to date (Revolution 162, 2004-02-10) 157 sources have been detected in a single ScW by ISGRI and/or JEM-X.  The positions of the detected sources are shown against the ISGRI exposure map in Fig.~\ref{exp}.  For each detected source, a database of the flux seen in each ScW, in either energy band of both instruments, is updated at the end of each revolution.  The summaries form a database, in text format, which can be queried quickly by a collection of simple scripts.  This allows the Scientist on Duty to easily investigate the flux history of any previously detected source, by producing a light curve, with fluxes binned per ScW.  A simple calibration can be made by scaling to the historic lightcurve for the Crab Nebula, as shown in the example in Fig.~\ref{vela}.  However, care must be taken to account for the differing calibration between older versions of the analysis and imaging software, as re-running the analysis for archival data each time a software update is made would require a prohibitive amount of computing time.

\section{Results from ISDC Operations}

Work at ISDC Operations is performed in close contact with the INTEGRAL Science Working Team (ISWT) and individual observers, depending on the proprietary rights of particular observations.  To date, a number of Scientific Circulars have arisen, at least in part, from the operational use of IQLA at the ISDC.  A selection of some of these, with reference to the relevant circulars, is shown below:

\begin{itemize}

\item 2004-02-19:  GX 339-4;  3--200 keV brightnening in JEM-X \& ISGRI -- ATEL 240

\item 2004-02-19:  4U 1724-307;  3--20 keV flux transition confirmed by JEM-X -- ATEL 241 

\item 2003-11-28:  Vela X-1;  Bright flare $>$  1 Crab  --   ATEL 211

\item 2003-08-26:  XTE J1739-302;   100 mCrab flare seen in 20--100 keV    --   ATEL 181 

\item 2003-04-23:  IGR J18325-0756;   New source seen by ISGRI                                --   ATEL 154

\item 2003-03-24:  XTE J1550-564;   20--100 keV flux $>$ 100 mCrab, triggered TOO  --    IAUC  8100   

\item 2003-03-06:  IGR J19140+098;   New source seen by ISGRI  --   IAUC 8088 

\item 2003-01-29:  IGR J16318-4848;   New source seen by ISGRI --   IAUC 8063

\end{itemize}

The ability of IQLA to automatically analyse INTEGRAL images as soon as they are available means that it is possible to report events such as those above within a few hours of their occurrence.  If necessary, the Scientist on Duty can perform a deeper analysis, with the standard Offline Science Analysis (OSA) from within the operational network, to confirm detections or make further measurements in one or more ScWs \citep{isdc}.  A list of Scientific Circulars arising from INTEGRAL observations is maintained at the ISDC at;

\texttt{\small{http://isdc.unige.ch/index.cgi?Science+circulars}}


Several new sources, that are bright enough to be seen in a single ScW
(whether persistent or transient) have also been reported by INTEGRAL. For
a summary of all new INTEGRAL sources, see;

\texttt{\small{http://isdc.unige.ch/$\sim$rodrigue/html/igrsources.html}}

\section{Conclusions}

INTEGRAL images are routinely monitored, every day of the year, by the
ISDC duty Scientist.  The primary tool for this is IQLA, which allows
the interactive analysis of ISGRI and JEM-X images, and to respond to
alerts for potentially interesting events issued automatically by the
system.  The ISDC IQLA software has also been saving a database of
source detections since revolution 81.  This database can be used to
show quick-look light curves of any previously detected source, which
is highly useful in investigating the flux history of a source
currently found to be active.  In the  future, it is envisaged that
light curves for all detected sources will be publically available.

To date, scientific operations of the ISDC have performed well, with
several new sources and transient events reported from the routine
monitoring of INTEGRAL data at the ISDC.  In comparison to previous
hard X-ray / soft gamma-ray instruments, INTEGRAL has a unique
property in that it is able to detect, confirm and report interesting
events within a few hours of their occurrence.  The monitoring of
INTEGRAL data, by the ISDC Scientists on Duty, is a significant part of this property.

\section*{Acknowledgements}

The hard work of the ISDC Operators, Software and Hardware Support staff, in continuing to ensure the smooth running of ISDC operations, is greatly appreciated by the authors.

\bibliographystyle{aa}

\end{document}